\documentclass[showpacs, preprintnumbers,superscriptaddressprl,floatfix,letterpaper,groupedaddress]{revtex4}
\usepackage{graphicx}
\usepackage{amsfonts}
\usepackage{amsmath, amsthm, amssymb}

\def \pslash {p\!\!\!/}				
\begin{document}

\title[Analytical formulas for elastic neutrino-fermion scattering cross sections]{Analytical formulas for elastic neutrino-fermion scattering cross sections}

\author{J. Linder}
\email{jacob.linder@ntnu.no}
\affiliation{Department of Physics, Norwegian University of
Science and Technology, N-7491 Trondheim, Norway}
\date{Received \today}
\begin{abstract}
\noindent In this paper, analytical formulas for elastic neutrino-fermion scattering cross sections are presented. 
These are evaluated in the limit where the neutrino mass is neglible compared to the incoming momentum. The fermion mass is taken into account, however, yielding a good approximation to realistic scattering reactions. The intended level of readership is for undergraduates/graduates in the fields of relativistic quantum mechanics and quantum field theory. In addition to a detailed and pedagogical derivation of the analytical expressions, numerical results are also provided for $f=e,n,p$.
\end{abstract}
\maketitle
\section{Introduction}
The motivation for writing a paper on analytical formulas for elastic neutrino-fermion scattering cross sections is that \textit{i)} their explicit derivation is hard to find in literature in general, \textit{ii)} their derivation provides very useful insight into how particle reactions are analyzed in the framework of quantum field theory and Feynman diagram analysis, and \textit{iii)} that some important equations and aspects of the analysis are often left out or sparsely treated in original textbooks. This paper should be of particular interest to students looking for a detailed review of how scattering reactions are dealt with, in addition to some important observations and techniques involved in this task. Textbooks such as Mandl\&Shaw \cite{qft} provide excellent introductory texts to the topic of quantum field theory, and equips the reader with many tools to handle scattering reactions. The "dirty" details of such calculations can be hard to find, however, and are highly valuable and appreciated by students trying to understand the bits and pieces of how one arrives at a given result.\\

This paper is organized as follows. In 
Sec. \ref{cs}, the concept of a reaction cross section is briefly reviewed to set the notation. Sec. \ref{EWneu} takes on 
neutrino interactions in electro-weak theory, which is the necessary framework for studying neutrino-fermion scattering. To set the notation, the relevant Feynman symbols are presented in Sec. \ref{feynman}. The 
general expressions for neutrino-fermion cross sections are derived in Sec. \ref{escatt}, and a summary is given in Sec. \ref{conclusion}. Unless specifically stated otherwise, natural units $\hbar=c=1$ will be used.

\section{Cross section}\label{cs}
The differential cross section d$\sigma$ says something about how incoming particles are scattered in space, and the likelihood of the scattering to occur at all. More precisely, since the interpretation of 
$\sigma$ is the number of particles scattered into a specific final state per unit time divided on the incident flux, 
the expression for d$\sigma$ is \\

\begin{equation}\label{diffcross1}
\mbox{d}\sigma = \frac{wV}{v_{\mbox{{\scriptsize rel}}}} \times \prod_f \frac{V \mbox{d}{\bf p}_f'}{(2\pi)^3},
\end{equation}

where $w$ is the transition probability from $|i\rangle$ to $|f\rangle$ per unit time, $v_{\mbox{{\scriptsize rel}}}$ is the relative velocity between incident and target particles, and $\prod_f \frac{V \mbox{{\scriptsize d}}{\bf p}_f'}{(2\pi)^3}$ is the 
number of states with momentum in the interval $({\bf p}_f', {\bf p}_f' + \mbox{d}{\bf p}_f')$. In the special case of
two incoming particles, the differential cross section reads

\begin{equation}\label{diffcross}
\mbox{d}\sigma = \frac{1}{4E_1E_2v_{\mbox{{\scriptsize rel}}}}(2\pi)^4 \delta^{(4)} \mbox{\Big (} \sum_f p'_f   - \sum_i p_i   \mbox{\Big )} \prod_f \frac{\mbox{d}{\bf p}_f'}{(2\pi)^32E'_f} |{\cal{M}}|^2,
\end{equation}

where $p_i$ are the 4-momenta of the incoming particles, $p'_f$ are the 4-momenta of the outgoing particles, and ${\cal{M}}$ 
is the Feynman amplitude of the process. More on this in Sec. \ref{EWneu}. \\

Calculations are often simplified when considering the process from the Center-of-Mass (CM) system. 
Here, Eq. (\ref{diffcross}) is reduced to

\begin{equation}\label{diffcrossCM}
\mbox{\huge (}\frac{\mbox{d}\sigma}{\mbox{d}\Omega}\mbox{\huge )}_{\mbox{\tiny CM}} = \frac{1}{64\pi^2(E_1 + E_2)^2}\frac{|\mbox{\bf p}_1'|}{|\mbox{\bf p}_1|} |{\cal{M}}|^2.
\end{equation}

For a detailed derivation of these quantities, consider for instance Ref. \cite{qft}.

\section{Neutrino interactions in electro-weak theory}\label{EWneu}

In the following, we shall not consider neutrino interaction with the electromagnetic field $A_\mu(x)$, since the electromagnetic transition moments of the neutrino are too weak for any significant coupling to photons. The upper limit for the neutrino magnetic moment is $\sim 10^{-10} \mu_{\mbox{\tiny B}}$, taken from the Particle Data Group \cite{pdd}. When dealing with neutrino interactions it is sufficient to consider the Lagrangian densities involving $W$ and $Z$ exchange for reactions involving $\nu_l$. We shall stick with the notation of Ref. \cite{qft}. Now, the interaction term reads 
\begin{align}\label{lneu}
{\cal{L}}_{\mbox{\scriptsize I}} &=  {\cal{L}}_{\mbox{\scriptsize I},W} + {\cal{L}}_{\mbox{\scriptsize I},Z} \notag \\
&= -\frac{g}{2\sqrt{2}} \bigl[ J^\mu_W(x)W_\mu^\dagger(x) + J^{\mu\dagger}_W(x)W_\mu(x) \bigr] -\frac{g}{\cos\theta_W} J^\mu_Z(x)Z^0_\mu(x),
\end{align}

where $J^\mu_W(x)$ and $J^\mu_Z$ are the charged and neutral currents 

\begin{align}\label{currents}
J^\mu_W(x) &= \sum_l \overline{l}(x)\gamma^\mu (1-\gamma_5) \nu_l(x) \notag \\
J^\mu_Z(x) &= \frac{1}{2} \sum_l \bigl[ \overline{\nu}_l \gamma^\mu (g_V^{\nu_l}-g_A^{\nu_l}\gamma_5)\nu_l + \overline{l}\gamma^\mu (g_V^l - g_A^l\gamma_5)l \bigr].
\end{align}

\begin{figure}[h!]
\centerline{\scalebox{0.6}{\includegraphics{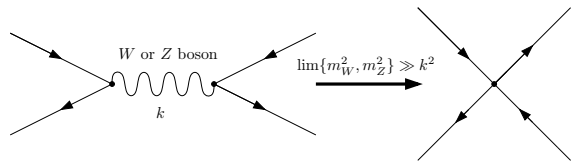}}}
\caption{\label{fig:trans} Effect of low-energy approximation with dominant $m_W$ and $m_Z$.}
\end{figure}

In Eq. (\ref{currents}), $l(x)$ and $\nu_l(x)$ represent the second quantized lepton fields. Furthermore, $g_V^i$ and $g_A^i$ are weak coupling constants, $\theta_W$ is the Weinberg angle, and $\gamma^\mu$ are the Dirac gamma matrices. $W_\mu(x)$ and $Z^0_\mu(x)$ describe the vector boson propagators found in Fig. \ref{fig:SEWfey}. Eq. (\ref{lneu}) is valid for all neutrino energies, but we would like to derive an expression for the low-energy limit which often occurs in realistic scattering experiments. More specifically, consider the limit of the external momenta being much smaller than the vector boson masses $m_W$ and $m_Z$. 
This means that

\begin{align}
\lim_{m_W^2\gg k^2} \Bigg[ \frac{-\mbox{i}(g_{\mu\nu} - k_\mu k_\nu / m_W^2)}{k^2 - m_W^2 + \mbox{i}\epsilon} \Bigg] = \frac{\mbox{i}g_{\mu\nu}}{m_W^2} \notag \\
\lim_{m_Z^2\gg k^2} \Bigg[ \frac{-\mbox{i}(g_{\mu\nu} - k_\mu k_\nu / m_Z^2)}{k^2 - m_Z^2 + \mbox{i}\epsilon} \Bigg] = \frac{\mbox{i}g_{\mu\nu}}{m_Z^2},
\end{align}

with $g_{\mu\nu} =$ diag(1,-1,-1,-1). Making these substitutions corresponds to the interaction transformation in Fig. \ref{fig:trans}, and thus provides us with the effective interaction Lagrangian density for neutrino interactions

\begin{align*}
{\cal{L}}^{\mbox{\scriptsize eff}}_{\mbox{\scriptsize I}} = {\cal{L}}^{\mbox{\scriptsize eff}}_{\mbox{\scriptsize I},W} + {\cal{L}}^{\mbox{\scriptsize eff}}_{\mbox{\scriptsize I},Z} = - \Bigl(\frac{g}{2\sqrt{2}}\Bigr)^2 \frac{1}{m_W^2} J^\mu_WJ_{W\mu}^\dagger - \Bigl(\frac{g}{\sqrt{2}\cos\theta_W}\Bigr)^2 \frac{1}{m_Z^2} J^\mu_ZJ_{Z\mu}^\dagger.
\end{align*}

The gauge boson masses are related through the Weinberg angle $\theta_W$ by $m_W/m_Z = |\cos\theta_W|$. $u$ and $d$ quarks 
can also be incorporated into this model, merely by adding extra terms to the currents in Eq. (\ref{currents}), 
thus producing

\begin{align}\label{modcurrents}
J^\mu_W &=  \sum_l \bigl[\overline{l} \gamma^\mu (1-\gamma_5) \nu_l + \overline{d}_{\theta} \gamma^\mu (1-\gamma_5)u \bigl] 
\notag \\ 
J^\mu_Z &= \frac{1}{2} \sum_l \bigl[ \overline{\nu}_l \gamma^\mu (g_V^{\nu_l}-g_A^{\nu_l}\gamma_5)\nu_l + 
\overline{l}\gamma^\mu (g_V^l - g_A^l\gamma_5)l  \notag \\
&\hspace{0.2in} + \overline{u}\gamma^\mu(g_V^u - g_A^u\gamma_5)u + \overline{d}\gamma^\mu(g_V^d - g_A^d\gamma_5)d\bigr],
\end{align}

where $d_{\theta} = d\cos{\theta_C} + s\sin{\theta_C}$. The Cabbibo angle $\theta_C$ has been experimentally determined to 
$\cos\theta_C \approx 0.98$, so we shall exclude the $s$ quark part from now on and set $d_{\theta} = d$. The introduction 
of quark terms enable us to describe processes such as $\nu_e + n \to p + e$. When adding the quark terms, we have used the 
general expression for the neutral current found in Ref. \cite{quigg}, namely

\begin{equation}
J^\mu_Z = \frac{1}{2} \sum_i \overline{\psi}_i \gamma^\mu \Bigl[I^3_i(1-\gamma_5) - 2Q_i\sin^2\theta_W\Bigr]\psi_i,
\end{equation}

where $i = (l,\nu_l,u,d)$, while $I^3_i$ is the belonging particle isospin and $Q_i$ is the particle charge in units of $e$. \\

Now, the Hamiltonian density ${\cal{H}}$ is related to its Lagrangian density ${\cal{L}}$ by ${\cal{H}} = \frac{\partial{\cal{L}}}{\partial\dot\phi^k}\dot\phi^k - {\cal{L}}$, 
with $\phi = \phi(x)$ denoting the various fields appearing in ${\cal{L}}$, i.e. $l(x), \nu_l(x)$ et.c. The second quantized expansion for Dirac fields reads

\begin{align}\label{lepfield}
l(x) = \sum_{s\mathbf{p}} \sqrt{\frac{1}{2VE_{\mathbf{p}}}} \Big[ a_{s}(\mathbf{p})u_s(\mathbf{p})\mbox{e}^{-\mbox{\scriptsize i}px} + b_{s}^\dag(\mathbf{p})v_s(\mathbf{p})\mbox{e}^{\mbox{\scriptsize i}px}\Big] \notag \\
\overline{l}(x) = \sum_{s\mathbf{p}} \sqrt{\frac{1}{2VE_{\mathbf{p}}}} \Big[ b_{s}(\mathbf{p})\overline{v}_s(\mathbf{p})\mbox{e}^{-\mbox{\scriptsize i}px} + a_{s}^\dag(\mathbf{p})\overline{u}_s(\mathbf{p})\mbox{e}^{\mbox{\scriptsize i}px}\Big],
\end{align}

and similar for $\nu_l(x), u(x), d(x)$. Here, $E_{\mathbf{p}}$ is the energy of the lepton with momentum $\mathbf{p}$, $\{a_{s}(\mathbf{p}),a^\dag_{s}(\mathbf{p})\}$ and $\{b_{s}(\mathbf{p}),b^\dag_{s}(\mathbf{p})\}$ are annihilation and creation operators for particles and antiparticles, respectively, while $\{u_s(\mathbf{p}),v_s(\mathbf{p})\}$ are Dirac spinors. Note that $\overline{l}(x) \equiv l^\dag(x)\gamma^0$. \\

Constructing the interaction Hamiltonian density for a process is a matter of straight-forward 
calculation, which is considerably simplified by making the observation that ${\cal{L}}$ does not contain any time 
derivatives of the fields, i.e. ${\cal{L}} = - {\cal{H}}$. We arrive at ${\cal{H}}^{\mbox{\scriptsize eff}} = {\cal{H}}^{\mbox{\scriptsize eff}}_{\mbox{\scriptsize I},W} + 
{\cal{H}}^{\mbox{\scriptsize eff}}_{\mbox{\scriptsize I},Z}$, where ${\cal{H}}^{\mbox{\scriptsize eff}}_{\mbox{\scriptsize I},W} = \frac{G_F}{\sqrt{2}} J_{W\mu} J^{\mu\dag}_W$, 
${\cal{H}}^{\mbox{\scriptsize eff}}_{\mbox{\scriptsize I},Z} = \frac{4G_F}{\sqrt{2}} J^\mu_Z J_{Z\mu}$. The Fermi constant is defined as $G_F = \sqrt{2}g^2/8m_W^2$. 

\section{Feynman rules}\label{feynman}

\begin{figure}[h!]
\centerline{\scalebox{1.0}{\includegraphics{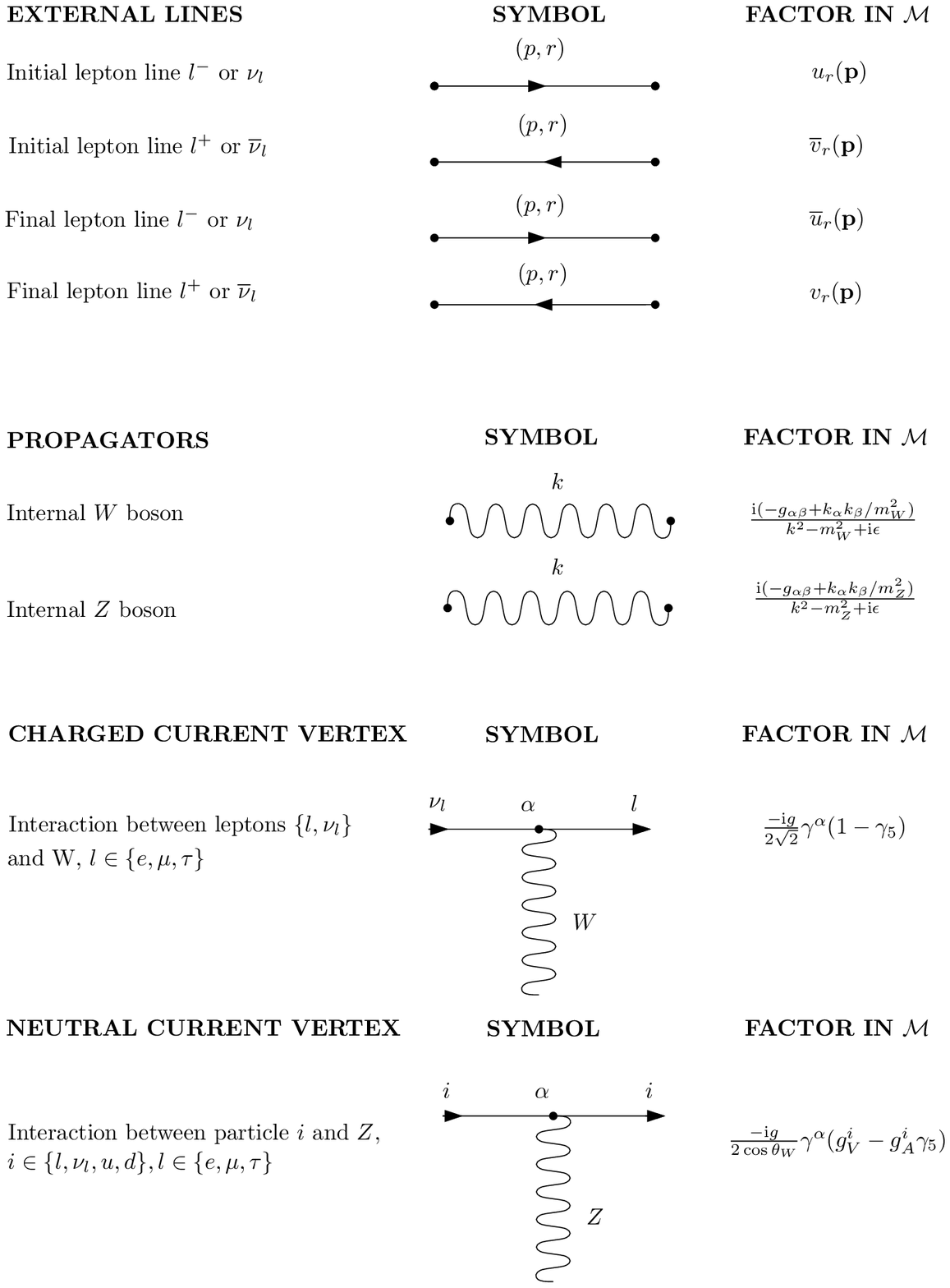}}}
\caption{\label{fig:SEWfey} Algebraic expressions for Feynman symbols in standard weak theory, relevant for our needs.}
\end{figure}

The Feynman diagram of a process is an indispensible tool for analysis of any particle reaction, in order to distuingish between 
different outcomes and to equip us with an elegant manner of calculating important physical quantities, such as the 
cross section. The Feynman amplitude can be read from the diagram with some training and quantifies the process. 
For our needs, we shall only give a short overview of the symbols and algebraic equivalents used in Feynman diagrams 
describing weak processes to set the notation. This is seen in Fig. \ref{fig:SEWfey}. 

\section{Neutrino-fermion scattering}\label{escatt}

In this section, we first give attention to neutral current neutrino-fermion scattering, i.e. mediated by the $Z^0$ boson. Then, we generalize to reactions mediated by the $W$ boson. Then it is time to say something about the coupling constants appearing in the expressions for the total cross section.

\subsection{Neutral current scattering}
\begin{figure}[h!]
\centerline{\scalebox{0.75}{\includegraphics{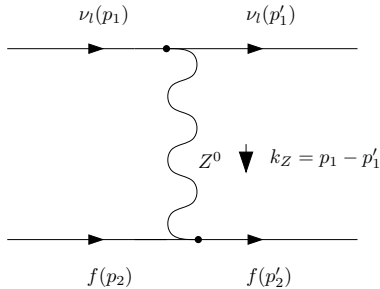}}}
\caption{\label{fig:f2} Lowest contributing order Feynman diagrams for $\nu_l + f \to \nu_l + f$.}
\end{figure}

To begin with, we shall consider $\nu_lf\mbox{ }(l\neq f)$ scattering. This process is mediated only by the $Z^0$ boson, if we
disregard the Higgs boson as a gauge particle. This is justified by the fact that the Feynman amplitude corresponding 
to the scattering process mediated by $H$ is of ${\cal{O}}(m_{\nu_l}m_{f}/m_H^2)$ compared to the $Z^0$ diagram. The 
only relevant lowest order Feynman diagram for $\nu_lf\mbox{ }(l\neq f)$ scattering is shown in Fig. \ref{fig:f2}.

The Feynman amplitude is given by

\begin{align*}
{\cal{M}} &= \overline{u}_{\nu_l}(1')[-\frac{\mbox{i}g}{2\cos\theta_W}\gamma^\alpha(g_V^{\nu_l}-g_A^{\nu_l}\gamma_5)]u_{\nu_{f}}(1)\mbox{i}\mbox{\Big (}\frac{-g_{\alpha\beta} + k_{Z\alpha} k_{Z\beta} / m_Z^2}{k_Z^2 - m_Z^2 + \mbox{i}\epsilon}\mbox{\Big )}  \\
&\times \mbox{ } \overline{u}_{f}(2')[-\frac{\mbox{i}g\gamma^\beta}{2\cos\theta_W}(g_V^{f} - g_A^{f}\gamma_5)]u_{f}(2,) \\
\end{align*}

Observe that we have shortened down $u_{s_i}({\bf p}_i)$ to $u(i)$. \\

These expressions can be simplified a great deal, when making some considerations. It is reasonable to expect $m_W^2 \gg k_W^2$ when using $m_W = 80.4$ GeV for the intermediate 
vector boson $W^\pm$. The same inequality goes for $m_Z = 91.2$ GeV and $m_H \geq 112 $ GeV. By applying this 
low-energy limit, we regain the effective propagator from Sec. \ref{EWneu}, arriving at 

\begin{eqnarray}\label{mll1}
{\cal{M}} = \frac{-2\mbox{i}G_F}{\sqrt{2}}\overline{u}_{\nu_l}(1')[\gamma^\alpha(g_V^{\nu_l}-g_A^{\nu_l}\gamma_5)]
u_{\nu_l}(1)\overline{u}_{f}(2')[\gamma_\alpha(g_V^{f} - g_A^{f}\gamma_5)]u_{f}(2).
\end{eqnarray}

We now pursue the absolute square of the Feynman amplitude, $|{\cal{M}}|^2$, which is part of the differential 
cross section from Sec. \ref{cs}. In a perfect world, the polarization, i.e. spins, of the reacting particles 
would be known. Unfortunately, this is the exception rather than the rule. If one does not know the initial and final 
spins of the neutrinos and electrons, this must be compensated for by averaging over the incoming spin states and adding 
the final states. This normally corresponds to an averaging factor 1/2 for each incoming particle. \\

Neutrinos, however, are special cases. Although they are massive, their mass is so small that it is ordinarily set to 
zero in the standard model. This means that there is only one helicity state for the neutrino, since it consequently 
travels with the speed of light and never can be overcome in any Lorentz frame. Thus, there is only one spin state to 
consider for the neutrinos which leads to the substitution
\begin{equation*}
|{\cal{M}}|^2 \to \frac{1}{2} \sum_{s_i, s'_f} |{\cal{M}}|^2 
\end{equation*}
in order to obtain the unpolarized cross-section. By exploiting the fact that $|{\cal{M}}|^2 = {\cal{M}}{\cal{M}}^*$ and 
$(\gamma^\mu)^\dagger = \gamma^0\gamma^\mu\gamma^0$, the problem is attacked. Inserting Eq. (\ref{mll1}) gives 

\begin{eqnarray*}
\frac{1}{2} \sum_{s_i, s'_f} |{\cal{M}}|^2  = \frac{1}{2} \sum_{s_i, s'_f} \mbox{\Bigg [} \frac{-2\mbox{i}G_F}{\sqrt{2}}
\overline{u}_{\nu_l}(1')\gamma^\alpha(g_V^{\nu_l}-g_A^{\nu_l}\gamma_5)u_{\nu_{f}}(1)\overline{u}_{f}(2')
\gamma_\alpha(g_V^{f} - g_A^{f}\gamma_5)u_f(2) \mbox{\Bigg ]} \\
\times \mbox{\Bigg [} \frac{2\mbox{i}G_F}{\sqrt{2}}\overline{u}_{f}(2)\gamma_\beta(g_V^{f} - g_A^{f}\gamma_5)
u_{f}(2')\overline{u}_{\nu_l}(1)\gamma^\beta(g_V^{\nu_l}-g_A^{\nu_l}\gamma_5)u_{\nu_l}(1')
\mbox{\Bigg ]}
\end{eqnarray*}

\begin{figure}[h!]
\centerline{\scalebox{0.6}{\includegraphics{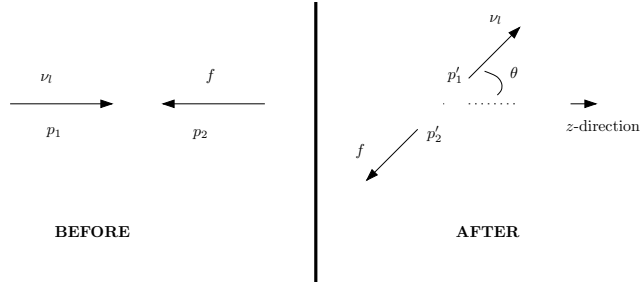}}}
\caption{\label{fig:cm} $\nu_lf\mbox{ } (l\neq f)$ scattering in the CM frame.}
\end{figure}

We stride on by recalling the completeness relations for the Dirac spinors $u$ and $v$
\begin{equation*} 
\sum_s u_{l\alpha}(p,s)\overline{u}_{l\beta}(p,s) = (\pslash + m_l)_{\alpha\beta}, \hspace{0.1in} \sum_s v_{l\alpha}(p,s)\overline{v}_{l\beta}(p,s) = (\pslash - m_l)_{\alpha\beta}.
\end{equation*}
Attaching indices on spinors and matrices provides us with
\begin{align}
\frac{1}{2} \sum_{s_i, s'_f} |{\cal{M}}|^2  = G_F^2 \sum_{s_i, s'_f} &\mbox{\Bigg [} \overline{u}_{\nu_l\kappa}(1')
\mbox{\Big [}\gamma^\alpha(g_V^{\nu_l}-g_A^{\nu_l}\gamma_5)\mbox{\Big ]}_{\kappa\lambda}u_{\nu_{l}\lambda}(1)\overline{u}_{f\gamma}(2')
\mbox{\Big [}\gamma_\alpha(g_V^{f} - g_A^{f}\gamma_5)\mbox{\Big ]}_{\gamma\delta}u_{f\delta}(2) \mbox{\Bigg ]} \notag \\
\times &\mbox{\Bigg [} \overline{u}_{f\epsilon}(2)\mbox{\Big [}\gamma_\beta(g_V^{f} - g_A^{f}\gamma_5)
\mbox{\Big ]}_{\epsilon\tau}u_{f\tau}(2')\overline{u}_{\nu_l\sigma}(1)\mbox{\Big [}\gamma^\beta(g_V^{\nu_l}-g_A^{\nu_l}\gamma_5)
\mbox{\Big ]}_{\sigma\rho}u_{\nu_l\rho}(1')\mbox{\Bigg ]} \notag \\
= G_F^2 \mbox{Tr} \mbox{\Bigg \{} &(\pslash_1' + m_{\nu_l})\gamma^\alpha(g_V^{\nu_l}-g_A^{\nu_l}\gamma_5)
(\pslash_1 + m_{\nu_l})\gamma^\beta(g_V^{\nu_l}-g_A^{\nu_l}\gamma_5) \mbox{\Bigg \}}  \mbox{\hspace{1.2in}}\notag \\
\times \mbox{Tr} \mbox{\Bigg \{} &(\pslash_2' + m_{f})\gamma_\alpha(g_V^{f} - g_A^{f}\gamma_5
)(\pslash_2 + m_{f})\gamma_\beta(g_V^{f} - g_A^{f}\gamma_5)\mbox{\Bigg \}}. 
\end{align}
We designate 
\begin{align}
\mbox{\textcircled{1}} &= \mbox{Tr} \mbox{\Bigg \{} (\pslash_1' + m_{\nu_l})\gamma^\alpha(g_V^{\nu_l}-g_A^{\nu_l}\gamma_5)
(\pslash_1 + m_{\nu_l})\gamma^\beta(g_V^{\nu_l}-g_A^{\nu_l}\gamma_5) \mbox{\Bigg \}} \notag \\
\mbox{\textcircled{2}} &= \mbox{Tr} \mbox{\Bigg \{} (\pslash_2' + m_{f})\gamma_\alpha(g_V^{f} - g_A^{f}\gamma_5
)(\pslash_2 + m_{f})\gamma_\beta(g_V^{f} - g_A^{f}\gamma_5)\mbox{\Bigg \}},
\end{align}
and consider each of the traces separately. Using the general relations for traces
\begin{align}\label{gammaprop}
&\mbox{Tr}\{A + B + ...\} = \mbox{Tr}\{A\} + \mbox{Tr}\{B\} + ... \notag \\
&\mbox{Tr}\{\gamma^5\} = \mbox{Tr}\{\gamma^5\gamma^\alpha\} = \mbox{Tr}\{\gamma^5\gamma^\alpha\gamma^\beta\} = \mbox{Tr}\{\gamma^5\gamma^\alpha\gamma^\beta\gamma^\gamma\} = 0 \notag \\
&\mbox{Tr}\{\gamma^5\gamma^\alpha\gamma^\beta\gamma^\gamma\gamma^\delta\} = -4\mbox{i}\epsilon^{\alpha\beta\gamma\delta} \notag \\
&\mbox{Tr}\{ \gamma^\alpha\gamma^\beta ... \gamma^\rho\gamma^\sigma\} = 0 \mbox{ if }\gamma^\alpha\gamma^\beta ... \gamma^\rho\gamma^\sigma\mbox{is an odd number of $\gamma$-matrices,} 
\end{align}
where $\epsilon^{\alpha\beta\gamma\delta}$ is a completely antisymmetric tensor, we obtain 
\begin{align}
\mbox{\textcircled{1}} =& 8p_{1\mu} ' p_{1\nu}\mbox{\Big [} \mbox{i}g_V^{\nu_l}g_A^{\nu_l}\epsilon^{\mu\alpha\nu\beta} + \frac{1}{2}[(g_V^{\nu_l})^2 + (g_A^{\nu_l})^2] (\eta^{\mu\alpha}\eta^{\nu\beta} + \eta^{\mu\beta}\eta^{\nu\alpha} - \eta^{\mu\nu}\eta^{\alpha\beta}) \mbox{\Big ]} \notag\\
\mbox{\textcircled{2}} =& 8p_2'^\sigma  p_2^\rho \mbox{\Big [} \mbox{i}g_V^{f}g_A^{f} \epsilon_{\sigma\alpha\rho\beta} + \frac{1}{2} [(g_V^{f})^2 + (g_A^{f})^2] ( \eta_{\sigma\beta}\eta_{\rho\alpha} + \eta_{\sigma\alpha}\eta_{\rho\beta} - \eta_{\sigma\rho}\eta_{\alpha\beta} ) + \frac{1}{2}m_f^2(g_V^2 - g_A^2)\eta_{\alpha\beta}\mbox{\Big ]}
\end{align}
in the limit $m_{\nu_l} \to 0$. From Eq. (\ref{diffcrossCM}) we see that the differential cross-section in the CM system 
must be
\begin{equation}\label{cross}
\Bigl(\frac{\mbox{d}\sigma}{\mbox{d}\Omega}\Bigl)_{\mbox{\tiny CM}} = \frac{X}{64\pi^2(E_1 + E_2)^2}\frac{|\mbox{\bf p}_1'|}{|\mbox{\bf p}_1|},
\end{equation}
where 
\begin{equation*}
X = \frac{1}{2} \sum_{s_i, s'_f} |{\cal{M}}|^2 = G_F^2\mbox{\big [}\mbox{\textcircled{1}} \times \mbox{\textcircled{2}}\mbox{\big ]}.
\end{equation*}
The product of the two traces is
\begin{align}\label{trprod}
\mbox{\textcircled{1}} \times \mbox{\textcircled{2}} &= 128g_V^{f}g_A^{f}g_V^{\nu_l}g_A^{\nu_l} \mbox{\big [} (p_1'p_2')(p_1p_2) -  (p_1'p_2)(p_1p_2') \mbox{\big ]} \notag\\
&+ 32 [(g_V^{f})^2 + (g_A^{f})^2][(g_V^{\nu_l})^2 + (g_A^{\nu_l})^2] \mbox{\big [} (p_1'p_2)(p_1p_2') + (p_1'p_2')(p_1p_2)  \notag\\
&- 32m_f^2[(g_V^{f})^2 - (g_A^{f})^2][(g_V^{\nu_l})^2 + (g_A^{\nu_l})^2](p_1'p_1)\mbox{\big ]}.
\end{align}
The situation now looks like Fig. \ref{fig:cm} in the CM frame with 4-momenta
\begin{align}
&p_1 = (E,0,0,E), \notag \\
&p_2 = (\sqrt{E^2 + m_{f}^2},0,0,-E), \notag \\
&p_1' = E(1,\cos\phi\sin\theta,\sin\phi\sin\theta,\cos\theta), \notag\\
&p_2' = (\sqrt{E^2 + m_{f}^2},-E\cos\phi\sin\theta,-E\sin\phi\sin\theta,-E\cos\theta),
\end{align}
where $E \equiv E_{\mbox{\tiny CM}}$ is the CM kinetic energy of the neutrino.

Since all 4-momenta now are known, insertion into Eq. (\ref{trprod}) gives

\begin{align}\label{trprod2}
\mbox{\textcircled{1}} \times \mbox{\textcircled{2}} &= 128g_V^{f}g_A^{f}g_V^{\nu_l}g_A^{\nu_l} \big[ (E\sqrt{E^2 + m_{f}^2} + E^2)^2 - (E\sqrt{E^2 + m_{f}^2} + E^2\cos\theta)^2 \big] \notag\\
&+ 32 [(g_V^{f})^2 + (g_A^{f})^2][(g_V^{\nu_l})^2 + (g_A^{\nu_l})^2] \big[ (E\sqrt{E^2 + m_{f}^2} + E^2)^2 + (E\sqrt{E^2 + m_{f}^2} + E^2\cos\theta)^2 \notag\\
&-32[(g_V^{f})^2 - (g_A^{f})^2][(g_V^{\nu_l})^2 + (g_A^{\nu_l})^2]E^2(1-\cos\theta) \big]. 
\end{align}

The total cross section for $\nu_lf\mbox{ } (l\neq f)$ scattering is now within our grasp, and reads

\begin{equation}\label{sigma1}
\sigma[\nu_lf\mbox{ } (l\neq f)] = \frac{G_F^2}{32\pi(E + \sqrt{E^2 + m_{f}^2})^2} \int^{\pi}_0 \mbox{\textcircled{1}} \times \mbox{\textcircled{2}} \sin\theta \mbox{d}\theta.
\end{equation}
From Eq. (\ref{trprod2}), it is clear that we must evaluate three types of integrals, namely
\begin{align}\label{eq:integrals}
\int^{\pi}_0 \sin\theta \mbox{ d}\theta = 2, \mbox{ } \int^{\pi}_0 \cos\theta\sin\theta \mbox{ d}\theta = 0, \mbox{ } \int^{\pi}_0 \cos^2\theta\sin\theta \mbox{ d}\theta = \frac{2}{3}.
\end{align}
Note that $s = (p_1+p_2)^2 = (E + \sqrt{E^2 + m_{f}^2})^2$. At this point, we make a little "cheat" and reveal the neutrino coupling constants $g_V^{\nu_l}$ and $g_A^{\nu_l}$ to be 1/2. We shall remedy this in Sec. \ref{coupling}. Inserting our integrals Eq. (\ref{eq:integrals}) reduces Eq. (\ref{sigma1}) to

\begin{align}\label{eq:crossnuf2}
\sigma[\nu_lf\mbox{ } (f\neq l)] = \frac{G_F^2(s-m_f^2)^2}{4\pi s} \Bigg[(g_V^f+g_A^f)^2 + (g_V^f - g_A^f)^2\Big[\frac{m_f^2}{s} + \frac{(s-m_f^2)^2}{3s^2}\Big] - [(g_V^f)^2-(g_A^f)^2]\frac{m_f^2}{s}  \Bigg]
\end{align}

The cross section for antineutrino-fermion scattering, i.e. $\overline{\nu}_lf (f\neq l)$, is obtained from the following
argumentation. In the limit $m_{\nu_l} \to 0, l=e,\mu,\tau$, neutrinos are always left-handed while antineutrinos 
will be right-handed. Thus, only a parity transformation ${\cal{P}}$ on the neutrino vertex part of the Feynman amplitude 
Eq. (\ref{mll1}) is required to 
obtain $\overline{\nu}_ll'\mbox{ } (l\neq l')$ scattering. Now, a scalar product of two axial vectors is invariant under a parity transformation, and so is 
the scalar product of the vector quantities as well. Recall that axial vectors $y^\mu=(y^0, {\bf y})$ transform as 
$y^\mu \stackrel{\cal{P}}{\longmapsto} \widetilde{y}^\mu =  (-y^0, {\bf y})$ under parity, while vector quantities 
$z^\mu = (z^0, {\bf z})$ transform as $z^\mu  \stackrel{\cal{P}}{\longmapsto} \widetilde{z}^\mu = (z^0, {\bf -z})$. 
Thus, it follows that a scalar product of one axial and vector quantity is not invariant under parity transformations, 
since
\begin{equation}
y^\mu z_\mu  \stackrel{\cal{P}}{\longmapsto} \widetilde{y}^\mu\widetilde{z}_\mu =  -y^0z_0 + {\bf yz} = -y^\mu z_\mu
\end{equation}
As a consequence, all mixed terms $g_V^{\nu_l}g_A^{\nu_l}$ will change sign. This means that the cross section for antineutrino-fermion scattering is found by making the substitutions

\begin{equation}
\sigma[\overline{\nu}_lf\mbox{ }(l\neq f)] = \lim_{g_V^{\nu_l}g_A^{\nu_l} \to (-g_V^{\nu_l}g_A^{\nu_l})}\sigma[\nu_lf\mbox{ } (l\neq f)].
\end{equation}

For $\overline{\nu}_l\overline{f}\mbox{ }(l\neq f)$ scattering, $g_V^{f}g_A^{f}$ will also change sign, yielding

\begin{equation}
\sigma[\overline{\nu}_l\overline{f}\mbox{ }(l\neq f)] = \sigma[\nu_lf\mbox{ } (l\neq f)].
\end{equation}

\subsection{Charged current scattering}
\begin{figure}[h!]
\centerline{\scalebox{0.75}{\includegraphics{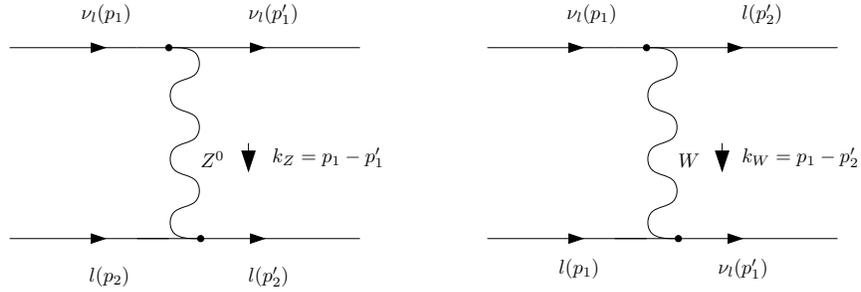}}}
\caption{\label{fig:3fey2} Lowest contributing order Feynman diagrams for $\nu_l + l \to \nu_l + l$.}
\end{figure}

Having derived the general expression for elastic neutrino-fermion scattering processes mediated by the $Z^0$ boson, we can now include the case $f=l$. This process is mediated by the $W$ boson in addition to $Z^0$, thus providing Feynman diagrams as shown in Fig. \ref{fig:3fey2}.

The total Feynman amplitude in the low-energy limit $k^2\ll\{m_Z^2,m_W^2\}$ is now ${\cal{M}} = {\cal{M}}_Z + {\cal{M}}_W$, where
\begin{align}
{\cal{M}}_Z &= \frac{-2\mbox{i}G_F}{\sqrt{2}}\overline{u}_{\nu_l}(1')[\gamma^\alpha(g_V^{\nu_l}-g_A^{\nu_l}\gamma_5)]
u_{\nu_l}(1)\overline{u}_{l}(2')[\gamma_\alpha(g_V^{l} - g_A^{l}\gamma_5)]u_{l}(2),\notag\\
{\cal{M}}_W &= \frac{-\mbox{i}G}{\sqrt{2}}\overline{u}_{l}(2')[\gamma^\alpha(1-\gamma_5)]u_{\nu_l}(1)\overline{u}_{\nu_l}(1')[\gamma_\alpha(1-\gamma_5)]u_l(2).
\end{align}

By using the Fierz identity 
\begin{equation}\label{fierz}
\overline{u}_1[\gamma^\alpha(1-\gamma_5)]u_2\overline{u}_3[\gamma_\alpha(1-\gamma_5)]u_4 = \overline{u}_1[\gamma^\alpha(1-\gamma_5)]u_4\overline{u}_3[\gamma_\alpha(1-\gamma_5)]u_2
\end{equation}
for arbitrary spinors $u_i$, ${\cal{M}}_W$ obtains the same sequence of spinors as ${\cal{M}}_Z$. In addition, 
we take advantadge of the fact that $g_V^{\nu_l} = g_A^{\nu_l} = 1/2$, $l=e,\mu,\tau$. This leaves us with

\begin{equation}\label{mll}
{\cal{M}} = \frac{-2\mbox{i}G}{\sqrt{2}}\overline{u}_{\nu_l}(1')\mbox{\Big [}\gamma^\alpha(g_V^{\nu_l}-g_A^{\nu_l}\gamma_5)\mbox{\Big ]}u_{\nu_l}(1)\overline{u}_l(2')\mbox{\Big [}\gamma_\alpha[(g_V^l+1) - (g_A^l+1)\gamma_5]\mbox{\Big ]}u_l(2).
\end{equation}

This is exactly the amplitude Eq. (\ref{mll1}) with the replacements $g_V^{f}\to g_V^{l}+1$, 
$g_A^{f}\to g_A^{l}+1$. Therefore, we are immediately able to write down the total cross section for 
elastic $\nu_ll$ scattering, namely

\begin{equation}
\sigma[\nu_ll] = \lim_{\stackrel{ g_V^{f}\to g_V^{l}+1 }{_{ g_A^{f}\to g_A^{l}+1 }}}\sigma[\nu_lf\mbox{ } (l\neq f)].
\end{equation}

\subsection{Coupling constants}\label{coupling}
The derivation of the coupling constants starts from the condition of U(1) and SU(2) gauge invariance for the standard electro-weak Lagrangian. Although an interesting exercise in itself, it is outside the scope of this paper, and we simply state the results in Tab. \ref{tab:coupling}. \\

\begin{table}[h!]
\centering{
	\caption{Relevant weak coupling constants for our needs with $l=\{e,\mu,\tau\}$ from Ref. \cite{peskin}.}
	\label{tab:coupling}
	\vspace{0.15in}
	\begin{tabular}{ccc}
	  	 \hline
	  	 \hline
		 {\bf Particle	}	\hspace{0.3in}		& \bf Axial coupling $g_A$	\hspace{0.3in}	& \bf Vector coupling $g_V$	\\
	  	 \hline
		$l$			\hspace{0.3in}		& $-1/2$		\hspace{0.3in}	& $2\sin^2\theta_W - 1/2$ \\
		$\nu_l$			\hspace{0.3in}		& $1/2$			\hspace{0.3in}	& $1/2$\\
		$u$			\hspace{0.3in}		& $1/2$			\hspace{0.3in}	& $1/2 - 4\sin^2\theta_W/3$ \\
		$d$			\hspace{0.3in}		& $-1/2$		\hspace{0.3in}	&$ 2\sin^2\theta_W/3 - 1/2$ \\
	  	 \hline
	  	 \hline
	\end{tabular}}
\end{table}

It should be mentioned that our factor $g_A^p = 2g_A^u + g_A^d$ is not exactly equal to 1/2. This is due to the axial current anomaly, which states that the conservation law for the axial current is broken. Ref. \cite{peskin} gives a proper treatment of this topic. Neutron beta-decay experiments have determined the value of axial vector coupling constant for protons to $g_A^p \simeq 0.635$ (see Ref. \cite{pdd}).

\section{Summary}\label{conclusion}
We have found that the elastic cross section for $\nu_lf$ scattering, where $l=e,\mu,\tau$ and $f\neq l$ is a fermion with vector coupling strength $g_V^f$ and axial coupling strength $g_A^f$, is given by\\

\begin{align}\label{eq:crossnuf2a}
\sigma[\nu_lf\mbox{ } (f\neq l)] = \frac{G_F^2(s-m_f^2)^2}{4\pi s} \Bigg[(g_V^f+g_A^f)^2 + (g_V^f - g_A^f)^2\Big[\frac{m_f^2}{s} + \frac{(s-m_f^2)^2}{3s^2}\Big] - [(g_V^f)^2-(g_A^f)^2]\frac{m_f^2}{s}\Bigg],
\end{align}

where $s = (E_{\mbox{\tiny CM}} + \sqrt{E_{\mbox{\tiny CM}}^2 + m_{f}^2})^2 = 2E_{\mbox{\tiny LAB}}m_f + m_f^2$. For charged current $\nu_ll$ scattering, the cross section is obtained through

\begin{equation}
\sigma[\nu_ll] = \lim_{\stackrel{ g_V^{f}\to g_V^{l}+1 }{_{ g_A^{f}\to g_A^{l}+1 }}}\sigma[\nu_lf\mbox{ } (f\neq l)].
\end{equation}

Two specific limits of Eq. (\ref{eq:crossnuf2a}) are of particular interest. By taking $m_f \to 0$, the resulting cross section corresponds to elastic neutrino scattering on light leptons. This approximation is very good for $f = \{e,\mu\}$. In this case, Eq. (\ref{eq:crossnuf2a}) reduces to

\begin{equation}
\lim_{m_f\to0}\sigma[\nu_lf\mbox{ } (f\neq l)] = \frac{G_F^2s}{3\pi} \Big[ g_V^2 + g_A^2 + g_Vg_A \Big],
\end{equation}

where we have dropped the supercript $f$ for the fermions. Another interesting scenario is the limit $E_{\mbox{\tiny LAB}}/m_f \ll 1$. The physical interpretation of such a limit is neutrino scattering on heavy fermions, {\it e.g.} protons, where the resulting recoil energy of the fermion is small. One finds that

\begin{equation}\label{eq:neupro}
\lim_{E_{\mbox{\tiny LAB}}/m_f \ll 1}\sigma[\nu_lf\mbox{ } (f\neq l)] = \frac{G_F^2E_{\mbox{\tiny LAB}}^2}{\pi} \Big[ g_V^2 + 3g_A^2 \Big].
\end{equation}

\begin{figure}[h!]
\centerline{\scalebox{0.49}{\includegraphics{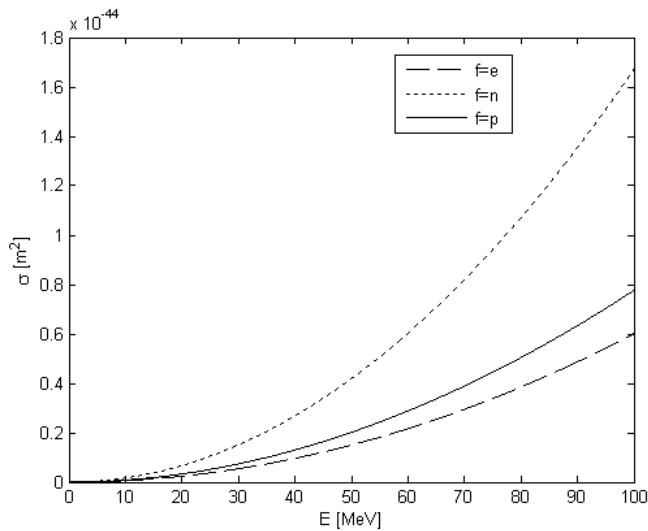}}}
\caption{\label{fig:crossenp} The elastic neutrino-fermion scattering cross section Eq. (\ref{eq:crossnuf2}) for $f=e,n,p$ as a function of the CM neutrino kinetic energy $E$.}
\end{figure}

\begin{figure}[h!]
\centerline{\scalebox{0.49}{\includegraphics{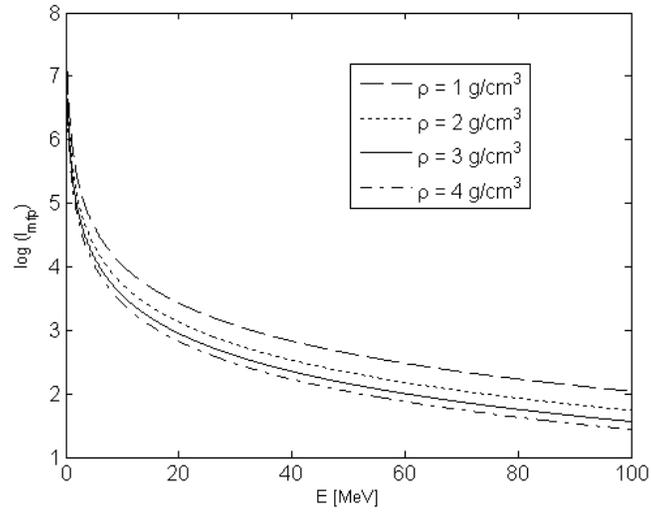}}}
\caption{\label{fig:length} Log-plot of mean free path length $l_{\mbox{\scriptsize mfp}}$ in meters as a function of CM neutrino kinetic energy $E$ for various mass densities.}
\end{figure}

This result agrees with theoretical predictions and experimental data (see {\it e.g.} Ref. \cite{proton1}). Fig. \ref{fig:crossenp} contains a plot of Eq. (\ref{eq:crossnuf2}) for the most common fermion scattering components on Earth, $f=e,n,p$. It is also instructive to consider the mean free path length $l_{\mbox{\scriptsize mfp}}$ for the neutrinos in a medium. This quantity is given as $l_{\mbox{\scriptsize mfp}} = \sum_f 1/(N_f\sigma_f)$, where $N_f$ is the number density of scattering component $f$. Taking $f=e,n,p$, the resulting mean free path length is shown in Fig. \ref{fig:length} for varying mass densities $\rho = \sum_f N_fm_f$, assuming an electrically neutral medium with $N_f \equiv N$. \\

\end{document}